\documentclass[preprint2]{aastex}

\shorttitle{On the connection between shape and stellar population in early-type galaxies}
\shortauthors{D'Onofrio et al.}

\begin{document}


\title{On the connection between shape and stellar population in early-type galaxies}


\author{M. D'Onofrio\altaffilmark{1} and T. Valentinuzzi\altaffilmark{1}}
\affil{Astronomy Department, Vicolo Osservatorio 3, I-35122 Padova, Italy}
\email{mauro.donofrio@unipd.it}
\author{G. Fasano\altaffilmark{2} and A. Moretti\altaffilmark{2} and D. Bettoni\altaffilmark{2} and B. Poggianti\altaffilmark{2} and B. Vulcani\altaffilmark{1,2} and J. Varela\altaffilmark{2}}
\affil{INAF/Astronomical Observatory of Padova, Vicolo Osservatorio 5, I-35122 Padova, Italy}
\author{J. Fritz\altaffilmark{3}}
\affil{Sterrenkundig Observatorium, University of Gent Krijgslaan 281 S9, B-9000 Gent, Belgium}
\author{A. Cava\altaffilmark{4}}
\affil{Instituto de Astrofisica de Canarias, via Lactea E38205, La Laguna (Tenerife), Spain}
\author{P. Kj{\ae}rgaard\altaffilmark{5}}
\affil{The Niels Bohr Institute for Astronomy Physics and Geophysics, Juliane Maries Vej 30, 2100, Copenhagen, Denmark}
\author{M. Moles\altaffilmark{6}}
\affil{Centro de Estudios de Fisica del Cosmos de Aragon (CEFCA), C/General Pizarro, 1, 44001 Teruel, Spain}
\author{W.J. Couch\altaffilmark{7}}
\affil{Center for Astrophysics and Supercomputing, Swinburne University of Technology, PO Box 218, Hawthorn Victoria 3122, Australia}
\and
\author{A. Dressler\altaffilmark{8}}
\affil{Observatories of the Carnegie Institution, 813 Santa Barbara Street, Pasadena, California 91101, USA}



\begin{abstract}
We report on the discovery of a relation between the stellar mass $M^*$ of
early-type galaxies (hereafter ETGs), their shape, as parametrized by
the Sersic index $n$, and their stellar mass-to-light ratio $M^*/L$. In a 3D
log space defined by these variables the ETGs populate a plane surface 
with small scatter. This relation tells us that galaxy shape and
stellar population are not independent physical variables, a result that
must be accounted for by theories of galaxy formation and evolution.
\end{abstract}




\section{Introduction}

It has long been known that the shape of galaxies is connected with
their stellar population properties. \citet{Morgan} based his galaxy
classification scheme of spiral galaxies on the relation between the central light
concentration and spectral characteristics \citep[see also][]{Morgan2}. 

ETGs have been considered a homogeneous class within the Hubble
sequence, manifesting a restricted range of shapes and a generic old
stellar population. Their internal structure is usually described as
differing by only a simple scale factor from galaxy to galaxy
(structural homology). Observational support for this interpretation
was based on the fact that all luminosity profiles appeared to follow
the de Vaucouleurs' law \citep{deVauc} $I(r)\sim r^{1/4}$. First,
\citet{Michard} and \citet{Schombert} noted that residuals of the
individual $r^{1/4}$ fits correlated with galaxy luminosity. Then, the
first breakthrough in the previously outlined scenario came more than
20 years ago through the papers by \citet{Cap87,deCdaC,Cap89,Burkert}. The final
proof of structural non-homology came when better fits of the
luminosity profiles were obtained using a Sersic' law $I(r)\sim
r^{1/n}$ \citep{Sersic}, where $n$ is a free index that defines the
shape of the light profile and correlates with the total luminosity of
the galaxy: the $n - L$ relation \citep{Caon,Young,Prugniel}.

It is also well known that ETGs share a 3D planar distribution with a
small scatter, the so-called Fundamental Plane
\citep[hereafter FP,][]{DjDa,Dressler}, defined by the effective
radius $R_e$, the
effective surface brightness $\langle\mu\rangle_e$, and the central
velocity dispersion $\sigma$. The observed tilt of the FP relative to the
virial theorem expectation, and the small scatter around the plane,
are still matters of debate. Many researchers believe that the tilt is
likely a consequence of stellar population effects which manifest
themselves through the observational evidence that low-mass ETGs tend
to be younger and to show lower stellar mass-to-light ratios with respect to
the massive ones \citep[the $M^*/L - L$ relation; see
e.g.][]{Faber87,Gerhard,Treu05}. Actually, the observations reveal that the
tilt is still substantial in the $K$ band \citep[$M_{tot}\sim L_K^{1.25\pm0.05}$, see
e.g.][]{Pahre}, where the effect of the stellar mass-to-light ratio variation
should be mild. Another possible explanation of the tilt is
linked to the amount and distribution of dark matter (DM). According
to \citet{Ciotti} the DM explanation requires a peculiar fine-tuning
to reproduce the observed properties of the FP. More recently
\citet{Tortora}, estimating the total $M/L$ from simple Jeans
dynamical models, found that an important role should be played by
DM, since its fraction within $R_e$ seems to be roughly constant for galaxies
fainter than $M_B \sim -20.5$, while it increases for brighter galaxies.
\citet{Padmanabhan} and \citet{Hyde} also found evidence that the
$M_{tot}/M^*$ ratio increases with mass, a fact that may imply a
dark-to-bright matter ratio increasing along the FP.

Another popular explanation of the FP tilt, alternative (or
complementary) to the previous ones, rests upon the already mentioned
evidence of structural non-homology of ETGs, parametrized by the
Sersic index $n$ \citep{Bertin,Trujillo}.

Here we report one of the results of a work in progress
\citep{Donofrio}, indicating that a connection between galaxy shape
and stellar population that was found by \citet{Morgan}
for spiral galaxies, is also present in ETGs. We suggest that such a
relationship is likely connected to the properties of the FP, in
particular with its scatter, and is therefore one of the fundamental
physical relations that underpins its origin.

\section{Our galaxy sample selection}


The present work is based on the data of the Wide-field Nearby
Galaxy-clusters Survey \citep[WINGS, see e.g.][]{Fasano,Varela,Cava,Valentinuzzi,Dono}.  
Our sample comes from the cross-match of the $V$ and $K$ band surface photometry.  Total
luminosities, effective radii and Sersic indices were
measured with the automatic software GASPHOT \citep{Pignatelli}.
The quality of the GASPHOT photometry
(magnitudes, radii, surface brightness and flattening) is discussed in
detail in several WINGS papers
\citep{Fasano,Varela,Cava,Valentinuzzi,Dono,Fritz,Fritz2}. 

{Galaxy masses have been derived by \citet{Fritz2} fitting the observed
(optical) spectra of the WINGS survey \citep{Cava} with theoretical
spectra, obtained summing up Simple Stellar Populations (SSP) spectra
of different ages, each one properly extinguished according to the
SSP's age itself \citep{Fritz}.  Details about the models and the
fitting procedure (evolutionary tracks, mass range and evolutionary
stages of stars, IMF, SFH, etc.) can be found in \citet{Fritz,Fritz2}. 
Here we just mention that the equivalent widths of the main lines and
the flux emitted in significant ranges of the spectral continuum have
been used to find the best fit of the observed spectra, at varying the
mass fractions of the SSPs, as well as the "selective extinction" and
the metallicity.  The masses obtained in this way have been found in
fair agreement with those obtained from our GASPHOT photometry,
following the prescriptions of \citet{Bell2}.

All galaxies were classified as early-types (pure ellipticals or S0s)
using MORPHOT, an automatic morphological classification package
purposely devised for the WINGS project
\citep{Fasano1,Fasano2}. 

Galaxy magnitudes, effective radii and Sersic indices have been found
in good agreement both with the SDSS data \citep{Valentinuzzi1} and
with the values we obtained (for comparison) using the GALFIT
\citep{Peng} and GIM2D \citep{Simard} packages.

From an initial sample of 2613 (848) galaxies with $V$ ($K$) band
data, we extracted a second "high quality sample" consisting of 240
ETGs with good photometry in both bands, well defined morphological
types, optimum Sersic light profile fits and robust stellar mass
measurements. The selected galaxies are all spectroscopically
confirmed cluster members.

\section{The $\log(M^*)-\log(n)-\log(M^*/L)$ relation}


The $n - L$ and $M^*/L - L$ scaling relations mentioned in the
introduction have been extensively discussed in the literature
\citep{Caon,Prugniel,Ciotti91,Busarello,Trujillo,GrahamColless,Bertin,Faber87,Gerhard,Treu05}.
They provide evidences that faint ETGs are characterized by small
values of $n$ and $M^*/L$ with respect to bright objects. The origin
of such behaviour is still a matter of debate.

In this letter we highlight the connection between $n$ (i.e. the shape
of the light profiles) and $M^*/L$ (i.e. the stellar populations) which
is hidden in the $M^*-n-M^*/L$ relation. The link between these two
fundamental variables that characterize ETGs was indeed poorly
analyzed in the literature, since the two variables do not correlate
directly.

\begin{figure}
\epsscale{.90}
\plotone{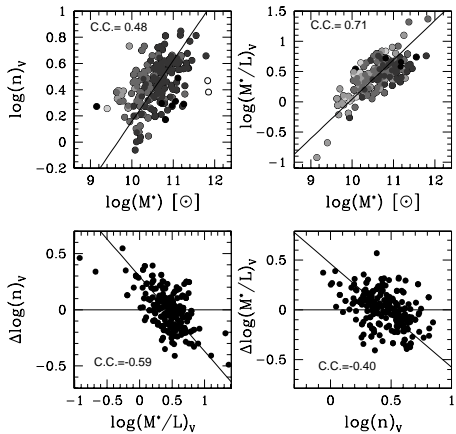}
\caption{({\it Upper panels}) The $\log(n)-\log(M^*)$ (left) and $\log(M^*/L)-\log(M^*)$ (right) relations. 
The gray scale of the dots mark low(white) to high(black) values of $\log(M^*/L)$ and $\log(n)$ respectively. 
The lines show the best fits. ({\it Lower panels}) The orthogonal residuals of the above relations 
plotted versus $\log(M^*/L)$ and $\log(n)$ respectively.\label{Fig1}}
\end{figure}

\begin{figure}
\epsscale{.95}
\plotone{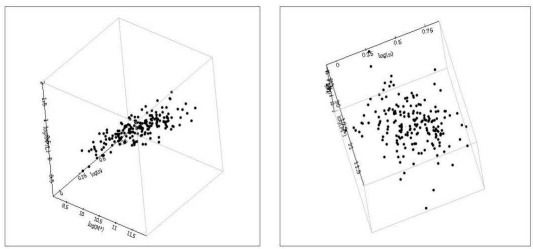}
\caption{The plane formed by ETGs in the 3D space defined by $\log(M^*)$, $\log(n)$ and 
$\log(M^*/L)$ seen along two different viewing angles.\label{Fig2}}
\end{figure}

The upper panels of Fig.~\ref{Fig1} show the $\log(n)-\log(M^*)$ and
$\log(M^*/L)-\log(M^*)$ relations for the "high quality sample". The
lines mark the bilinear least-square fits to the data obtained using
the SLOPE program \citep{Feigelson}. The lower panels of the same
figure show the orthogonal residuals of the two relations versus
$\log(M^*/L)$ and $\log(n)$, respectively. Note that the same trend of
the residuals is observed when the classical least-square fit is
adopted and the residuals are evaluated along the Y axis.  It is clear
from the plots that the residuals of the $\log(M^*/L)-\log(M^*)$ and
$\log(n)-\log(M^*)$ relations depend, respectively, upon $\log(n)$ and
$\log(M^*/L)$. These two variables are independent of each other, but
are connected through their residuals. The same mass-to-light ratio is
achieved by galaxies with a low/high mass when their Sersic index
($n$) is high/low, while the same value of $n$ is achieved by galaxies
of low/high masses when $M^*/L$ is high/low. Despite the well known
difficulties in measuring the Sersic index (variance $\Delta n\sim
1$), the observed trends reveal a clear interplay between galaxy
shape, mass and mass-to-light ratio. This behavior of the residuals
implies that ETGs are distributed on a plane in a 3D space whose axes
are $\log(M^*)$, $\log(n)$ and $\log(M^*/L)$ (see
Fig.~\ref{Fig2}). This relationship is observed in both $V$ and $K$ bands
and for both the "high quality" and the total samples. In particular,
for the "high quality" sample the equations of the planes which minimize
the sum of the distance from the fit are:

\begin{eqnarray}
\log(M^*) & = & 1.12(\pm0.22)\log(M^*/L)_V \nonumber \\
          &   & + 2.19(\pm0.23)\log(n_V) \nonumber \\
          &   & + 9.14(\pm0.11)  \\
\log(M^*) & = & 1.77(\pm0.55)\log(M^*/L)_K \nonumber \\
          &   & + 2.11(\pm0.26)\log(n_K) \nonumber \\
          &   & + 9.70(\pm0.14),
\end{eqnarray}

for the $V$ ($rms=0.10$) and $K$ ($rms=0.11$) bands, respectively. 
The Principal Component Analysis (PCA) confirms that we have a planar
distribution. The first two eigenvectors for the $V$ band are:

\begin{eqnarray}
\xi & = & 0.67\log(M^*)+0.59\log(n) \nonumber \\
    &   & +0.45\log(M^*/L) \\
\eta& = & 0.08\log(M^*)+0.54\log(n) \nonumber \\
    &   & -0.83\log(M^*/L).
\end{eqnarray}

They are built with the three variables, and together support up to
$\sim90\%$ of the total variance (respectively $\sim60\%$ and
$\sim30\%$). The same is true for the $K$ band with small differences
in the PCA coefficients.

As far as we know, the $n-M^*$ relation for ETGs is explicitly
presented here for the first time, at least for ETGs. Actually,
\citet{Dutton} published the $n-M^*$ relation for disk galaxies in the
redshift range $0.005<z<0.05$ using the data of the New York
University Value Added Catalog \citep[NYU-VAGC,][]{Blanton}. Their
slope is a bit lower than our, likely because of the limited range of
values of $n$ for disk objects (from 0.2 to 0.6 in log
scale). However, remarkably enough, their Fig. 5 clearly shows that
the residuals with respect to a straight line fit depend on colors,
i.e. on the $M^*/L$ ratio. This suggest that the $M^*-n-M^*/L$
relation might be valid over the whole Hubble sequence.

We can compare our $n-L$ and $n-R_e$
relations with that of \citet{Caon}, based on the Virgo and Fornax
ETGs. For the $V$ band we have:
$\log(n)=0.60(\pm0.05)\log(L)-5.74(\pm0.3)$ with a correlation coefficient $C.C.=0.51$ and
$\log(R_e)=0.73(\pm0.06)\log(n)+0.02(\pm0.03)$ with a $C.C.=0.54$.
Both relations are in agreement (within $3\sigma$) with the previous 
determinations by \citet{Caon}. Actually, in the case of the $n-L$ 
relation our slope is larger than in \citet[][0.6 vs. 0.45]{Caon}. However, 
we must consider that the slope value depends on the adopted fitting strategy. 
In fact, using a simple least-square fit, we get the relation
$\log(n)=0.30(\pm0.05)\log(L)-2.42(\pm0.3)$, quite different from that 
obtained when a bilinear
LSQ is used.  The data of \citet{Caon} are based on 40 ETGs against
the 240 of the WINGS database. Since the errors are approximately of
the same order for both variables, we believe that the bilinear LSQ
fit is more appropriate for determining the correct slope of the
relation. Therefore we estimate that the true slope of the $n-L$
relation is in the interval $0.4\div0.6$.

In the $V$ band our $M^*/L-M^*$ relation has a slope of $\sim0.6$
(again the slope depends on the fitting strategy adopted, varying from
$\sim0.45$ (for the classic LSQ fit) to $\sim0.6$ (for the bilinear
fit)), while the $M^*/L-L$ relation has a slope of $\sim0.03$ (with a
C.C.$\sim0.03$), in agreement with the recent determination by
\citet{Tortora}. Note that, given the slope of our $M^*/L-M^*$
relation (0.6), some trivial algebra would imply that $M^*/L \sim
L^2$, largely at variance with the formal value we found (2.0
vs. 0.03). However, it is easy to recognize the origin of this
apparent discrepancy in the lack of correlation between $M^*/L$ and
$L$ (C.C.$\sim0.03$), which makes practically undefined the slope of
the best fitting straight line. Such lack of correlation is also the
reason why there is a 3D planar distribution of ETGs in the
$M^*-n-M^*/L$ log space, while it is not present in the corresponding
$L-n-M^*/L$ space.

Since we were not able to find in the literature any estimate of the
slope of the $M^*/L-M^*$ relation, we tried to get an 'indirect' check
of our value combining the slopes of the $\sigma - M^*$ and $M^*/L -
\sigma$ relations.  Our data give $\sigma \sim M^{*\,0.26\div0.38}$,
depending on the adopted fitting tool (these values are within the
range of slopes given by \citealt{Graves}), and $M^*/L \sim
\sigma^{0.4\div1.3}$.  The combination of these equations provides
$M^*/L \sim M^{*\,0.1\div0.6}$, in agreement with our 'direct'
finding. In any case, it is worth stressing again that the existence
of the 3D $M^* - n - M^*/L$ relation does not depend on the final
slopes adopted for the $M^*-n$ and $M^*/L-M^*$ relations.

\section{Consequences for theories of galaxy formation and evolution}

Before the discovery of the FP \citep{DjDa,Dressler}, the $L-\sigma$,
$\sigma - R_e$ and $\langle\mu\rangle_e - R_e$ relations
\citep{FJ,Kormendy} were well known scaling relations of ETGs. When
\citet{Terlevich} found that the residuals of the Faber-Jackson
relation were connected with the effective surface brightness
$\langle\mu\rangle_e$, the existence of a 3D relationship between
$\sigma-R_e-\langle\mu\rangle_e$ (i.e. the FP) was soon
revealed. Similarly, we report here on the existence of a
plane preferentially shared by ETGs in the 3D space defined by the
quantities $M^*$, $n$ and $M^*/L$, which are already known to be
linked by mutual independent correlations. We define this plane
and claim that its very existence implies that stellar populations and
shapes of ETGs are not independent quantities. Why this occurs is presently
unknown, but this mutual dependence recalls the correlation found by
\citet{Morgan} for spiral galaxies and could be an important
ingredient for theories of galaxy formation and evolution. A 'natural'
guess one can make is that such a relation is connected to
the FP of ETGs, i.e. with the process of virialization that produced,
after the gas collapse phase, the final stellar structure in dynamical
equilibrium. How can this finding be reconciled with the
hierarchical merging scenario of galaxy formation where ETGs are the
final output of a series of merging events?  How is this relationship
connected with the well known correlations of galaxy mass and
mass-to-light ratio with age and metallicity? How it is connected with
the FP? We are presently investigating the implications of our
finding. Preliminary analyses based on FP simulations in which the
Sersic indices of the mock galaxies randomly share the observed
distribution, or (alternatively) are forced to obey (with the proper
scatter) the relation (1),
suggest that it might act as a sort of 'fine-tuning' effect that keeps
small the scatter around the FP \citep{Donofrio}.


\acknowledgments

Benedetta Vulcani and Bianca Maria Poggianti acknowledge the financial
support from the ASI contract I/016/07/0. We also thank the anonymous referee
for his help in improving this work and for bringing our attention toward
the paper of \citet{Dutton}.

\clearpage



\begin{thebibliography}{}
\bibitem[Bell \& de Jong(2002)]{Bell2} Bell, E.F., \& de Jong, R.S. 2001, \apj\ 550, 212
\bibitem[Bertelli et al.(1994)]{Bertelli} Bertelli, G., Bressan, A., Chiosi, C., Fagotto, F., \& Nasi, E. 1994, A\&AS, 106, 275
\bibitem[Bertin, Ciotti \& Del Principe(2002)]{Bertin} Bertin G., Ciotti L., \& Del Principe M. 2002, \aap\ 386, 149
\bibitem[Blanton et al.(2005)]{Blanton} Blanton M.R. et al. 2005, \aj\ 129, 2562
\bibitem[Burkert(1993)]{Burkert} Burkert A. 1993, \aap\ 278, 23
\bibitem[Busarello et al.(1997)]{Busarello} Busarello G., Capaccioli M., Longo G., \& Puddu E. 1997, in: The Second Stromlo Symposium ``The nature of Elliptical Galaxies'', ASP Conference Series, 166, 184
\bibitem[Caon et al.(1993)]{Caon} Caon N., Capaccioli M., \& D'Onofrio M. 1993, \mnras\ 265, 1013
\bibitem[Capaccioli(1987)]{Cap87} Capaccioli M. 1987, The Structure and Dynamics of Elliptical Galaxies,  P.T. De Zeew ed. (Reidel, Dordrecht), 127, 47
\bibitem[Capaccioli(1989)]{Cap89}  Capaccioli M. 1989, The World of Galaxies, H.G. Corwin, L. Bottinelli eds. (Springer-Verlag, Berlin), 208
\bibitem[Cava et al.(2009)]{Cava} Cava A. et al. 2009, \aap\ 495, 707
\bibitem[Ciotti(1991)]{Ciotti91} Ciotti L. 1991, \aap\ 249, 99
\bibitem[Ciotti, Lanzoni \& Renzini(1996)]{Ciotti} Ciotti L., Lanzoni B., \& Renzini A. 1996, \mnras\ 282, 1
\bibitem[de Carvalho \& da Costa(1988)]{deCdaC} de Carvalho R.R., \& da Costa L.N. 1988, \apjs\ 68, 173
\bibitem[de Vaucouleurs(1948)]{deVauc} de Vaucouleurs G. 1948, Ann. d'Astrophys. 11, 247
\bibitem[Djorgovski \& Davis(1987)]{DjDa} Djorgovski S., \& Davis M. 1987, \apj\ 313, 59
\bibitem[D'Onofrio et al.(2008)]{Dono} D'Onofrio M. et al. 2008, \apj\ 685, 875
\bibitem[D'Onofrio et al.(2010)]{Donofrio} D'Onofrio M. et al. 2010, \mnras\ in preparation
\bibitem[Dressler et al.(1987)]{Dressler} Dressler A. et al. 1987, \apj\ 313, 42 
\bibitem[Dutton(2009)]{Dutton} Dutton A.A. 2009, \mnras\ 396, 121
\bibitem[Faber et al.(1987)]{Faber87} Faber S.M., Dressler A., Davies R., Burstein D., \& Lynden-Bell D. 1987, in: Nearly normal galaxies: From the Planck time to the present; Proceedings of the Eighth Santa Cruz Summer Workshop in Astronomy and Astrophysics, Santa Cruz, CA, July 21-Aug. 1, 1986 (A88-18401 05-90). New York, Springer-Verlag, 1987, p. 175-183 
\bibitem[Faber \& Jackson(1976)]{FJ} Faber S.M., Jackson R.E. 1976, \apj\ 204, 668  
\bibitem[Fasano \& Vanzella(2007)]{Fasano1} Fasano G., \& Vanzella E. 2007, From Stars to Galaxies: Building the Pieces to Build Up the Universe". ASP Conference Series, 374, 495
\bibitem[Fasano et al.(2006)]{Fasano} Fasano G. et al. 2006, \aap\ 445, 805
\bibitem[Fasano et al.(2010)]{Fasano2} Fasano G. et al. 2010, \aap\ in preparation 
\bibitem[Feigelson \& Babu(1992)]{Feigelson} Feigelson E.D., \& Babu G.J. 1992, \apj\ 397, 55
\bibitem[Fritz et al.(2007)]{Fritz} Fritz J. et al. 2007, \aap\ 470, 137
\bibitem[Fritz et al.(2010)]{Fritz2} Fritz J. et al. 2010, \aap\ accepted [arXiv:1010.2214] 
\bibitem[Gerhard et al.(2001)]{Gerhard} Gerhard O., Kronawitter A., Saglia R.P., \& Bender R. 2001, \aj\ 121, 1936
\bibitem[Graham \& Colless(1997)]{GrahamColless} Graham A., \& Colless M. 1997, \mnras\ 287, 221
\bibitem[Graves \& Faber(2010)]{Graves} Graves G.J., \& Faber S.M. 2010, arXiv:1005.0014v1
\bibitem[Hyde \& Bernardi(2009)]{Hyde} Hyde J.B., \& Bernardi M. 2009, \mnras\ 396, 1171
\bibitem[Kormendy(1977)]{Kormendy} Kormendy J. 1977, \apj 218, 333
\bibitem[Michard(1985)]{Michard} Michard R. 1985, A\&AS 59, 205
\bibitem[Morgan(1958)]{Morgan} Morgan W.W. 1958 \pasp\ 70, 364
\bibitem[Morgan \& Mayall(1957)]{Morgan2} Morgan W.W., \& Mayall N.U. 1957, PASP 69, 291
\bibitem[Padmanabhan et al.(2004)]{Padmanabhan} Padmanabhan N., et al. 2004, New Astronomy, 9, 329
\bibitem[Pahre et al.(1998)]{Pahre} Pahre M.A., De Carvalho R.R., \& Djorgovski S.G. 1998, \aj\ 116, 1606
\bibitem[Peng et al.(2002)]{Peng} Peng C.Y. et al. 2002, \aj\ 124, 266 
\bibitem[Pignatelli, Fasano \& Cassata(2006)]{Pignatelli} Pignatelli E., Fasano G., \& Cassata P. 2006, \aap\ 446, 373
\bibitem[Prugniel \& Simien(1997)]{Prugniel} Prugniel Ph., \& Simien F. 1997, \aap\ 321, 111
\bibitem[Salpeter(1955)]{Salpeter} Salpeter, E.E. 1955, \apj\ 121, 161
\bibitem[Schombert(1986)]{Schombert} Schombert J.M. 1986, \apjs\ 60, 603
\bibitem[Sersic(1968)]{Sersic} Sersic J.L. 1968, Atlas de Galaxias Australes, Observatorio Astronomico de Cordoba
\bibitem[Simard et al.(2002)]{Simard} Simard L., et al. 2002, \apjs\ 142, 1 
\bibitem[Terlevich et al.(1981)]{Terlevich} Terlevich R., et al. 1981, \mnras\ 196, 381
\bibitem[Tortora et al.(2009)]{Tortora} Tortora C., et al. 2009, \mnras\ 396, 1132
\bibitem[Treu et al.(2005)]{Treu05} Treu T., Ellis R.K., Liao T.X., \& van Dokkum P.G. 2005, \apj\ 622, L5
\bibitem[Trujillo, Burkert, \& Bell(2004)]{Trujillo} Trujillo I., Burkert A., \& Bell E.F. 2004, \apj\ 600, 39
\bibitem[Valentinuzzi et al.(2009)]{Valentinuzzi} Valentinuzzi T. et al. 2009, \aap\ 501, 851
\bibitem[Valentinuzzi et al.(2010)]{Valentinuzzi1} Valentinuzzi T. et al. 2010, \apj\ 721, L19
\bibitem[Varela et al.(2009)]{Varela} Varela J. et al. 2009, \aap\ 497, 667
\bibitem[Young \& Currie(1994)]{Young} Young C.K., \& Currie M.J. 1994, \mnras\ 268, 11
\end{thebibliography}
\end{document}